\documentstyle[twoside,fleqn,espcrc2,epsfig]{article}

 \newcommand{\be}{\begin{equation}}
 \newcommand{\ee}{\end{equation}}
 \newcommand{\ba}{\begin{eqnarray}}
 \newcommand{\ea}{\end{eqnarray}}
 \newcommand{\ifig}[1]{\mbox{\epsfig{file=#1,height=70mm}}}
 \makeatletter
 \@addtoreset{equation}{section}
 \makeatother
 \renewcommand{\theequation}{\thesection.\arabic{equation}}
 \def\appendix #1#2
 {
  \section*{Appendix #1: #2}
  \renewcommand{\theequation}{#1.\arabic{equation}}
 }

 \def\Schr{Sch\-r\"o\-din\-ger}

 \def\tanh{\hbox{\rm tanh}}
 \def\coth{\hbox{\rm coth}}

 \def\d{\partial}
 \def\h{\hat}

 \def\real{{\vrule height 1.6ex
            width 0.05em depth 0ex \kern -0.06em {\rm R}}}
 \def\entry#1#2{\vbox{\hbox to 112truept{\hrulefill}\break%
                \hbox{\vrule\vbox to 28truept{%
                \vfill%
                \hbox to 112truept{\hfill\quad\small #1\quad\hfill}\break%
                \vfill%
                \hbox to 112truept{\hfill\quad\small #2\quad\hfill}%
                \break\vfill%
                \hbox to 112truept{\hrulefill}}\vrule}}}%
 \def\arrwv#1#2{\vbox to 45truept{\vfill%
               \hbox to 112truept{\hfill
               \put(42,20){\small#1}
               \put(56,40){\vector(0,-1){40}}
               \put(62,20){\small#2}\hfill}
               \vfill}}
 \def\arrwhdd#1#2{\vbox to 28truept{\vfill
                \hbox to 92truept{\hfill
                \put(44,2){\small#1}
                \put(4,14){\vector(1,0){86}}
                \put(44,18){\small#2}\hfill}
                \vfill}}
 \def\cross#1#2{\vbox to 45truept{\vfill%
                \hbox to 112truept{\hfill
                \put(0,32){\small#1}
                \put(14,42){\vector(2,-1){86}}
                \put(102,32){\small#2}
                \put(14,0){\vector(2,1){86}}\hfill}
                \vfill}}

\newcommand{\AmS}{{\protect\the\textfont2
  A\kern-.1667em\lower.5ex\hbox{M}\kern-.125emS}}

\title{Recent developments in quantum string cosmology\thanks{Presented by
Carlo Ungarelli}} 

\author{M. Cavagli\`a\address{Departamento de Fisica,
Universidade da Beira Interior\\ 
R.\ Marqu{\^e}s d'{\'A}vila e Bolama, 6200 Covilh{\~a}, Portugal} 
and C. Ungarelli\address{School of Computer
Science and Mathematics, University of Portsmouth, Mercantile House,
Hampshire Terrace, Portsmouth P01 2EG, UK}}

\begin{document}

\begin{abstract}
In this talk we discuss the quantisation of a class of string cosmology  models
characterised by scale factor duality invariance.  The amplitudes for the full
set of classically allowed  and forbidden transitions  are computed by applying
the  reduced phase space and path integral methods. In particular,  the path
integral calculation clarifies the meaning of the  instanton-like behaviour of
the transition amplitudes that has been first  pointed out in previous
investigations. 
\end{abstract}

\maketitle

\section{Introduction}
One of the main problems of `pre-big bang' string cosmology~\cite{GV0,GV1,GV2} 
is the understanding of the mechanism responsible  for the transition
(`graceful exit') from the  inflationary  `pre-big bang' (PRBB) phase to a
deflationary  `post-big bang' phase (POBB) with decreasing curvature  typical
of the standard cosmological scenario.  Necessarily, the graceful exit involves
a high-curvature, strong coupling,  regime where higher derivatives and string
loops terms must be  taken into account. It has been shown~\cite{Grex} that for
any choice of the  (local) dilaton potential no cosmological solutions that
connect smoothly the  PRBB and POBB phases do exist. As a consequence, at the
classical level  higher order corrections cannot be `simulated' by any
realistic dilaton  potential. 

At the quantum level the dilaton potential may induce the transition from the 
PRBB phase to the POBB phase. In this context, using the standard Dirac  method
of quantisation based on the Wheeler-De Witt equation, a  number of
minisuperspace models have been investigated in the
literature~\cite{GV,GVM,Mah}. The result of these investigations is a finite,
non-zero,  transition probability PRBB $\rightarrow$ POBB with a typical 
`instanton-like' dependence ($\sim\,\exp\{-1/g^2\}$) on the string coupling 
constant.

In this talk we present a refined analysis of the quantisation of string
cosmological models by reconsidering the minisuperspace models previously
investigated~\cite{GV,GVM,Mah}. We will consider a class of string inspired
models that are exactly integrable and apply the standard techniques for the
canonical quantisation  of constrained systems.
\section{Classical theory}
We consider the string inspired model in d+1 dimensions described by the action
\be
S =\int\,\frac{d^{{\rm d}+1}x}{2\lambda_s^{{\rm d}-1}}\,\sqrt{|g|}\,e^{-\phi}\,\left(R+\partial_{\mu}\phi\partial^{\mu}\phi
-V\right),
\label{Act1}
\ee
where $\phi$ is the dilaton field, $\lambda_s=(\alpha^{\prime})^{1/2}$ is the
fundamental string length parameter, and $V=V(g_{\mu\nu},\phi)$ is a potential 
term. We deal with isotropic, spatially flat, 
cosmological backgrounds with finite volume spatial sections. For this
class of backgrounds the action (\ref{Act1}) reads
\be
S=\int\,dt\,
{\lambda_s\over 2}\, \left({1\over\mu}(\dot{\Theta}^2-\dot{\Phi}^2)-
\mu\,e^{-2\Phi}V\right)\,,\label{new-act}
\ee
where we have used the metric parametrisation  $g_{\mu\nu}={\rm diag}
\left(-\mu^2(t)e^{-2\Phi}, a^2(t)\delta_{ij}\right)$, $i,j=1,..$d,
$a= \exp[\Theta (t)/\sqrt{{\rm d}}]$. $\Phi=\phi- \log\,\int\,d^{\rm
d}x/\lambda_s^{\rm d}-\sqrt{{\rm d}}\,\Theta$ is the `shifted' dilaton field.
We shall restrict our attention to models which exhibit scale factor duality
invariance~\cite{GV0} (in this case the potential term depends only on the 
shifted dilaton $\Phi$). Moreover, we shall consider potentials of the form
$V(\Phi)=\lambda\,e^{-2\Phi(q-1)}$, where $\lambda>0$ is a dimension-two
quantity (in natural units) and $q$ is a  dimensionless parameter. (This class
of potentials has been first discussed  in~\cite{Mah}.) It is also convenient
to use the canonical form for the action 
\be
S=\int dt\,\left[\dot\Theta \Pi_\Theta+\dot\Phi \Pi_\Phi-
{\cal H}\right]\,,{\cal H}=\mu(t)H\,,
\label{action-can}
\ee
where the Hamiltonian constraint $H$ reads 
\be
H={1\over 2\lambda_s}
\left(\Pi_\Theta^2-\Pi_\Phi^2+\lambda_s^2 V(\Phi)\,e^{-2\Phi}\right)\,.
\label{H}
\ee
For $q\neq 0$ the explicit solution of the equations  of motion has been
derived and discussed in~\cite{CU99}.  The expanding and contracting
backgrounds are identified by the value $\Pi_{\Theta}=k>0$ and
$\Pi_{\Theta}=k<0$, respectively ($k=0$ corresponds to  the flat
(d+1)-dimensional Minkowski space). Moreover, for  $q\leq 1$ we have two
distinct branches corresponding to PRBB $(+)$ and POBB $(-)$ states, that are
identified by  negative and positive values of $\Pi_{\Phi}$, respectively. 
\section{Quantum theory}
The class of models introduced in the previous section is described by a
time-reparametrisation invariant Hamiltonian system with two degrees of 
freedom. Thanks to the integrability properties
of this class of models, the standard techniques of quantisation of constrained
systems can be applied straightforwardly. In particular, since the
constraint $H$ is of the form $H=H_\Theta(\Theta)+H_\Phi(\Phi)$ the time
parameter can be defined by a single degree of freedom. 
Since we are interested in the calculation of the quantum transition 
probability from a (expanding) PRBB phase to a (expanding) POBB phase, 
it is natural to use the $(\Theta,\Pi_\Theta)$ degree of freedom to define 
the time of the system and fix the gauge. 
In this case the eigenstates of the effective Hamiltonian are 
identified by a continuous quantum number $k$ corresponding to the classical 
value of $\Pi_\Theta$. Wave functions that describe expanding (contracting) 
solutions are eigenstates of the effective Hamiltonian with $k>0$ ($k<0$).

Let us perform the canonical transformation 
$\Sigma=\lambda_s\Theta/\Pi_\Theta$, $\Pi_\Sigma=
\Pi_\Theta^2/2\lambda_s$~\cite{CD}. Since  $\Sigma$ is canonically  conjugate
to $H$ it defines a global time parameter. Thus   the gauge fixing identity can
be chosen as $F(\Sigma;t)\equiv \Sigma+t=0$, fixing the Lagrange multiplier as
$\mu=-1$. The  gauge-fixed action reads
\be
S_{\rm eff}=\int_{t_1}^{t_2} dt\,\left[\dot\Phi\,\Pi_\Phi\,-\,H_{\rm 
eff}(\Phi,\Pi_\Phi)\right]\,,
\label{action-eff-qsc}
\ee
where the effective Hamiltonian is
\be 
H_{\rm eff}(\Phi,\Pi_\Phi)={1\over 
2\lambda_s}\left(\Pi_\Phi^2-\lambda\lambda_s^2 
e^{-2q\Phi}\right)\,.\label{H-eff-qsc}
\ee
The system described by the effective Hamiltonian (\ref{H-eff-qsc}) is free of
gauge degrees of freedom and its quantisation can be performed using  the
standard techniques. We shall now discuss the  reduced phase space and path
integral quantisation procedures.  \subsection{Reduced phase space
quantisation}  The reduced phase space is described by a single degree of
freedom with  canonical coordinates $(\Phi\in\real,\Pi_\Phi\in\real)$.  The
\Schr\ equation reads
\be
-i{\d~\over\d t}\,\Psi=\, {1\over 2\lambda_s}\left[{\d^2~\over 
\d\Phi^2}+\lambda\lambda_s^2 e^{-2q\Phi}\right]\,\Psi\,.
\label{schroedinger-qsc}
\ee
The general solution of~(\ref{schroedinger-qsc})  can be written in terms of
the solutions  $\psi_{k,q}(\Phi)$  of the stationary \Schr\ equation $\h 
H_{\rm eff}\psi=E\psi$ with energy $E=k^2/2\lambda_s$
\be
\psi_{k,q}(z)=A_1(k,q)J_{i\nu}(z)+A_2(k,q)Y_{i\nu}(z)\,,
\label{sol-gen}
\ee
where $J_{i\nu}(z)$ and $Y_{i\nu}(z)$ are the Bessel functions of the first and
second kind of index $i\nu=i|k/q|$ and argument
$z=\sqrt{\lambda}\lambda_s\exp(-q\Phi)/|q|$. In particular, two sets of
orthonormal  stationary solutions can be chosen  as the stationary wave
functions corresponding to expanding PRBB and POBB phases, either in
perturbative regime or in the strong coupling regime. For $q>0$, the
normalised PRBB(+) and POBB (-) wave functions in the  weak coupling (W) and
strong coupling (S) regimes are
\ba
&&\psi^{\rm (\pm)}_W={1\over\sqrt{2\coth(\pi\nu)}}\left[\sqrt{\tanh(\pi\nu/2)}
\chi^{(1)}_{\nu}\mp \right .\nonumber \\
&&\left .i\sqrt{\coth(\pi\nu/2)}\chi^{(2)}_{\nu}\right]\,,
\label{wave-weak-pre}\\
&&\psi^{\rm (\pm)}_S={1\over\sqrt{2}}\left[\chi^{(1)}_{\nu}\mp
i\chi^{(2)}_{\nu}\right]\,,\label{wave-strong-prpo}
\ea
where $\chi^{(1)}_{\nu}(z)$ and $\chi^{(2)}_{\nu}(z)$ are linear combinations 
of the Hankel functions $H^{(1,2)}_{i\nu}(z)$.  
Using the two sets of wave functions (\ref{wave-weak-pre}) and 
(\ref{wave-strong-prpo}) it is possible to compute the amplitudes that 
correspond to different transitions (a detailed analysis of those
amplitudes can be found in~\cite{CU99}).
In particular, the transition probabilities that correspond to classically 
allowed transitions read 
\be
P_{S,W}^{(+,+)}=P_{S,W}^{(-,-)}={1\over 1+e^{-2\pi k/q}}\,.
\label{prob-prew-pres}
\ee
Transitions between PRBB weak (strong) coupling regime and POBB strong (weak)
coupling regime are classically forbidden;  the relative transition
probabilities are given by
\be
P_{W,S}^{(+,-)}=P_{S,W}^{(+,-)}={e^{-2\pi k/q}\over 1+e^{-2\pi k/q}}\,.
\label{prob-prew-poss}
\ee
The last and most interesting result is the probability of transition from the 
PRBB phase in the weak coupling regime to the POBB phase in the weak coupling 
regime
\be
P_{W,W}^{(+,-)}=4{e^{-2\pi k/q}\over 
\left(1+e^{-2\pi k/q}\right)^2}\,.\label{prob-prew-postw}
\ee
For $q=1$, the semiclassical limit ($k\gg 1$) of (\ref{prob-prew-postw})
coincides, apart from a normalisation  factor, with the
`reflection-coefficient' of~\cite{GV,GVM}. However,  the result
of~\cite{GV,GVM} should be considered as a  ratio between two different
transition probabilities rather  than a transition probability by itself.
Precisely,  the reflection-coefficient defined in~\cite{GV,GVM} is  $R\equiv
P_{S,W}^{(-,+)}/P_{S,W}^{(+,+)}=e^{-2\pi  k/q}$. 

These results show that the probabilities of classically 
forbidden transitions can be expressed, in the semiclassical limit, as power 
series of $e^{-2\pi k/|q|}$. Following~\cite{GV,GVM}, we find
\be
\exp(-2\pi k/|q|)=
\exp\left(-\sqrt{4d}\,\pi\Omega_s\over |q|g^2_s\,\lambda^d_s\right)\,,
\label{ist0}
\ee
where $\Omega_s$ is the proper spatial volume and $g_s=e^{\phi_s/2}$ is the 
value of the string coupling when the Hubble parameter is   equal to
$1/\lambda_s$. The `istanton-like'  behaviour of (\ref{ist0}) shows that the
probabilities of  classically forbidden transitions are peaked in the strong
coupling regime -- as it was first pointed out in~\cite{GV,GVM} --  where all
powers of $e^{-2\pi k/|q|}$ have to be taken into account.  The occurrence of
this istanton-like behaviour will be discussed  in the next subsection. 
\subsection{Path integral quantisation}
The string cosmology model that we are considering can also be quantised using
the functional approach. Applying the path integral formalism,  we compute  the
probability $P_{W,W}^{(+,-)}=|A_{W,W}^{(+,-)}|^2$  in the semiclassical
limit~\cite{CU99}.  The starting point of the functional approach is the path
integral in the  reduced space 
\be
I=\int_{\Phi(t_1)}^{\Phi(t_2)}\,{\cal 
D}\Phi\,\exp\left(i\int_{t_1}^{t_2}dt\,{\cal L}_{\rm 
eff}[\Phi,\dot\Phi]\right)\,,\label{path-gf-lagr-qsc}
\label{pathin}
\ee
where the effective Lagrangian can be obtained from the effective
Hamiltonian~(\ref{H-eff-qsc}). The transition amplitude $A_{W,W}^{(+,-)}$ is
defined by (\ref{pathin})  where the integral is evaluated on all paths that
satisfy the boundary conditions $z(-\infty)=\infty$, $z(\infty)=\infty$.
(Recall that $z=\sqrt{\lambda}\lambda_s\exp(-q\Phi)/|q|$.)  Let us consider the
analytical continuation of the variable $z$ to the complex  plane. The
effective Lagrangian is analytical in any point of  the complex plane  $({\rm
Re}(z),{\rm Im}(z))$ save for $z=0$. Classically, the transition from  the weak
coupling PRBB phase to the weak coupling POBB phase would correspond  to the
trajectory starting at $z=+\infty$, going left along the real axis (PRBB phase,
$\dot\Phi>0$), reaching the origin,  and finally going right along the real
axis to $z=+\infty$ (POBB phase, $\dot\Phi<0$). Clearly, since the Lagrangian
is singular in $z=0$ a classical continuous and differentiable solution does
not exist.
%
\begin{figure}[htb]
\vspace{9pt}
\ifig{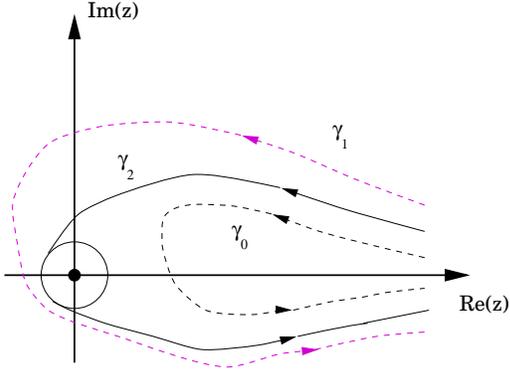}
\caption{Contours of integration in the complex $z$-plane.}
\label{figdue}
\end{figure}
%
Now consider generic analytical trajectories in the complex plane that start at
$Re(z)=\infty$, $Im(z)>0$, and end at $Re(z)=\infty$,  $Im(z)<0$ (see
Fig.\ref{figdue}). We can divide this class of trajectories in three
(topologically) distinct categories:  Trajectories that do not cross the
imaginary axis,  (curve $\gamma_0$ in Fig.\ref{figdue}); trajectories  that
cross twice the imaginary axis, (curve   $\gamma_1$ in Fig. \ref{figdue});
trajectories that cross $2n$  times ($n=2,3...$) the imaginary axis, (curve
$\gamma_n$ in Fig. \ref{figdue} for $n=2$). It is straightforward to see that 
the trajectories of type $\gamma_0$ do not describe transitions from PRBB to
POBB phases. A trajectory of type $\gamma_1$ describes a transition from the
weak coupling PRBB phase to the weak coupling  POBB phase. It can be deformed
continuously into a classical solution except in a small region in the strong
coupling limit, where the singularity of the classical solution is avoided by
the analytical continuation in the complex plane. The path integral evaluated
on this trajectory gives the leading contribution  to the semiclassical
approximation of the transition amplitude $A_{W,W}^{(+,-)}$. Trajectories of
type $\gamma_n$ (with $n\,>\,1$) give  contributes of higher order. In
particular, any trajectory that circles $z=0$ can be considered as an
`$n$-instanton'  solution (with no well-defined signature)  labelled by a
winding number $n$ that corresponds to the number of times that the trajectory
wraps around the singularity in $z=0$. In  the semiclassical limit, the
transition amplitude $P_{W,W}^{(+,-)}$  is given by the path integral
(\ref{path-gf-lagr-qsc})  evaluated on the class of  $n$-instanton solutions.
For the class of 1-instanton trajectories the amplitude is given by 
\be
A_{W,W}^{(+,-)(1)}=C_1 e^{-\pi k/q}\,,\label{prob-path}
\ee
where $C_1$ is a normalisation factor. 
The square of the semiclassical one-instanton 
amplitude (\ref{prob-path}) approximates  the 
(exact) result~(\ref{prob-prew-postw}) for large values of $k$. 
This proves the consistency of the reduced phase space and path 
integral quantisation methods. The contribution 
of the $n$-instanton ($n\,>\,1$) to the transition amplitude 
$A_{W,W}^{(+,-)}$ is
\be
A_{W,W}^{(+,-)(n)}=C_n e^{-\pi nk/q}\,.\label{prob-semicl-n}
\ee
Hence, $n$-instanton terms give higher order contributions in the large-$k$ 
expansion. Equations (\ref{prob-path}) and (\ref{prob-semicl-n}) show that the 
instanton-like dependence (\ref{ist0}) of the  amplitudes that correspond to
classically forbidden transitions can be traced  back to the existence, in the
semiclassical regime, of trajectories that  connect smoothly the PRBB and POBB
phases. 

\end{document}